\documentclass[10pt,conference]{IEEEtran}

\usepackage[usenames, x11names*]{xcolor}
\usepackage[T1]{fontenc}
\usepackage[utf8]{inputenc}
\usepackage[english]{babel}
\usepackage{csquotes}

\usepackage[pdftex]{graphicx}
\usepackage{mathtools,amssymb,amsfonts,bm, amsthm}
\makeatletter
\def\thm@space@setup{%
  \thm@preskip=1.5\topsep
  \thm@postskip=\thm@preskip % or whatever, if you don't want them to be equal
}
\makeatother
\usepackage[bookmarks=false]{hyperref}

\newcommand*\colourcheck[1]{%
	\expandafter\newcommand\csname #1check\endcsname{\textcolor{#1}{\ensuremath{\checkmark}}}%
}
\colourcheck{green}
\usepackage{caption}
\usepackage{subcaption}
\usepackage{booktabs}
\usepackage{url}
\usepackage[bookmarks=false]{hyperref}
\usepackage{float}
\usepackage{dblfloatfix}
\usepackage{csvsimple}

\usepackage{tikz}
\usetikzlibrary{quantikz,arrows.meta,backgrounds,calc,chains,matrix,positioning,shapes,shapes.geometric,	shapes.arrows, decorations.pathmorphing, decorations.pathreplacing}
\tikzset{%
	font={\footnotesize},
	vertex/.style={draw,circle,inner sep=0pt,minimum width=0.5cm,minimum height=0.5cm,font=\small},
	terminal/.style={draw,regular polygon,regular polygon sides=4,inner sep=0pt,minimum width=0.5cm,minimum height=0.5cm,font=\small, scale=1.0},
	zeroterm/.style={below,inner sep=0pt,font=\small, scale=0.3}
}

\usepackage{stmaryrd}
\usepackage[binary-units=true, decimalsymbol=comma, output-decimal-marker={.}]{siunitx}
\usepackage[textsize=tiny]{todonotes}
\usepackage{nicematrix}
\usepackage{nicefrac}

\usepackage{ifthen}
\usepackage{etoolbox}

\usepackage[]{flushend}

\usepackage[style=ieee, maxnames=6, minnames=1, date=year, doi=false,isbn=false,backend=biber, sortcites]{biblatex}
\makeatletter
\patchcmd{\maketitle}{\@copyrightspace}{}{}{}
\makeatother

\newtheorem{example}{Example}
\newcommand{\qop}[1]{\ensuremath{\mathit{#1}}}
\DeclarePairedDelimiterX{\abs}[1]{\lvert}{\rvert}{\ifblank{#1}{{}\cdot{}}{#1}}

\IEEEoverridecommandlockouts

\begin{document}

\title{Verifying Results of the IBM Qiskit \\Quantum Circuit Compilation Flow}
\author{
\IEEEauthorblockN{Lukas Burgholzer\IEEEauthorrefmark{1}\hspace{2.2cm}Rudy Raymond\IEEEauthorrefmark{2}\hspace{2.2cm}Robert Wille\IEEEauthorrefmark{1}\IEEEauthorrefmark{3}}
\IEEEauthorblockA{\IEEEauthorrefmark{1}Institute for Integrated Circuits, Johannes Kepler University Linz, Austria}
\IEEEauthorblockA{\IEEEauthorrefmark{2}IBM Research -- Tokyo, Japan}
\IEEEauthorblockA{\IEEEauthorrefmark{3}Software Competence Center Hagenberg GmbH (SCCH), Hagenberg, Austria}
\IEEEauthorblockN{lukas.burgholzer@jku.at\hspace{1cm}rudyhar@jp.ibm.com\hspace{1cm}robert.wille@jku.at}
\IEEEauthorblockN{\url{https://iic.jku.at/eda/research/quantum/}}
}
\date{}

\maketitle              % typeset the header of the contribution

\begin{abstract}Realizing a conceptual quantum algorithm on an actual physical device necessitates the algorithm's quantum circuit description to undergo certain transformations in order to adhere to all constraints imposed by the hardware. 
In this regard, the individual high-level circuit components are first synthesized to the supported low-level gate-set of the quantum computer, before being mapped to the target's architecture---utilizing several optimizations in order to improve the compilation result.
Specialized tools for this complex task exist, e.g., IBM's Qiskit, Google's Cirq, Microsoft's QDK, or Rigetti's Forest.
However, to date, the circuits resulting from these tools are hardly verified, which is mainly due to the immense complexity of checking if two quantum circuits indeed realize the same functionality.
In this paper, we 
propose an efficient scheme for \emph{quantum circuit equivalence checking}---specialized for verifying results of the IBM Qiskit quantum circuit compilation flow.
To this end, we combine characteristics unique to quantum computing, e.g., its inherent reversibility, and certain knowledge about the compilation flow %knowledge about the results to be verified 
into a dedicated equivalence checking strategy.
Experimental evaluations confirm that the proposed scheme allows to verify even large circuit instances with tens of thousands of operations within seconds or even less,  whereas state-of-the-art techniques frequently \mbox{time-out} or require substantially more runtime.
A corresponding open source implementation of the proposed method is publicly available at \url{https://github.com/iic-jku/qcec}.
\end{abstract}

\section{Introduction}

%Quantum computing is currently transitioning from being an emergent technology to becoming an established one, as actual devices reach feasibility.
Quantum computing has gained considerable momentum over the past years, as actual quantum computers are reaching feasibility and more and more algorithms for potential applications are discovered.
%We currently find ourselves in the NISQ era of quantum computing~\cite{preskillQuantumComputingNISQ2018}, in which actual devices reach feasibility and first stakeholders claim to have achieved quantum supremacy~\cite{aruteQuantumSupremacyUsing2019}.
Similar to the conventional realm, a conceptual quantum algorithm 
needs to be \emph{compiled} to a representation that conforms to all restrictions imposed by the device it shall be executed on.
To this end, fast-evolving compilation flows such as those of IBM's Qiskit~\cite{aleksandrowiczQiskitOpensourceFramework2019}, Google's Cirq~\cite{CirqPythonFramework}, Microsoft's QDK~\cite{QuantumDevelopmentKit}, or Rigetti's Forest~\cite{ForestSDK} are available.

Naturally, it is of utmost importance that the results of such compilation flows are correct, i.e., that %checking whether 
the compiled quantum circuit still realizes the originally intended functionality.
This motivates the development of methods for verification or, more precisely, equivalence checking.
Conceptually, equivalence checking is simple, since each quantum circuit~$G$ realizes a unitary transformation~$U$ and comparing two quantum circuits~$G$ and~$G^\prime$ boils down to comparing the corresponding 
unitary matrices~$U$ and~$U^\prime$, respectively.
However, since these matrices~$U$ and~$U^\prime$ are exponentially large, many existing approaches to this problem (e.g.,~\cite{viamontesCheckingEquivalenceQuantum2007, yamashitaFastEquivalencecheckingQuantum2010, smithQuantumComputationalCompiler2019,niemannEquivalenceCheckingMultilevel2014,wangXQDDbasedVerificationMethod2008}) remain unsatisfactory and can hardly be employed on a large scale. 
%\todo{ich habe hier jetzt die ansätze, die compiler selbst verifizieren nicht utner gebrachrt (mir war es irgendwie wichtiger, auf das U vs $U'$ einzugehen und dann wollte ich den roten faden nicht verlieren; mittlerweile frage ich mich eh, ob wir das nicht vllt ganz raus geben wollen und von anfang an gleich auf EC setzen (also auch in section iii raus?). was meinst du?}

%However, due to the immense complexity of the underlying problem and/or lacking adaptability of respective methods, existing solutions~\cite{hietalaVerifiedOptimizerQuantum2019, shiContractbasedVerificationRealistic2019, smithQuantumComputationalCompiler2019, wangXQDDbasedVerificationMethod2008, willeEquivalenceCheckingReversible2009, yamashitaFastEquivalencecheckingQuantum2010, burgholzerAdvancedEquivalenceChecking2020} remain unsatisfactory and can hardly be employed on a large scale.

Recently, a promising solution to address this problem has been proposed in~\cite{burgholzerAdvancedEquivalenceChecking2020}. 
There, the use of dedicated \mbox{data-structures}, in particular decision diagrams~\cite{niemannQMDDsEfficientQuantum2016, wangXQDDbasedVerificationMethod2008, zulehnerHowEfficientlyHandle2019}, has been proposed which allow for a non-exponential representation of matrices in some (albeit not all) cases. 
%\todo{see sreces}
%welcher satz besser (auskommentierte version oder derzeitig sichtbare?)
%To further improve upon that, the observation is exploited that, assuming two circuits~$G$ and~$G'$ are indeed equivalent, the cascade composed of one circuit with the inverse of the other should eventually yield the identity function, i.e., $G^{\prime -1}\cdot G = \mathbb{I}$. 
To further improve upon that, it is exploited that $G^{\prime -1}\cdot G = \mathbb{I}$ holds if two circuits~$G$ and~$G^\prime$ are indeed equivalent, i.e., the cascade composed of one circuit with the inverse of the other should eventually yield the identity function.  
Because of that, checking whether $G$ and $G^\prime$ are in fact equivalent can be conducted by starting with the identity matrix (which can be represented in linear rather than exponential space) and, then, applying operations of~$G$ and~$G^\prime$ in a particular order until all operations have been applied. 
If all those operations can be applied so that the respective intermediate computations remain as close to the identity as possible, substantial improvements can be achieved~\cite{burgholzerAdvancedEquivalenceChecking2020}.
However, how to determine the ``perfect'' order of applications, i.e., whether to apply operations from~$G$ or from~$G^\prime$ % in a certain step 
in order to keep %which keeps %in order to keep
the respective intermediate representation close to the identity, remains an open problem.\footnote{This is discussed and illustrated in more detail later in Section~\ref{sec:motivation}.}

In this paper, we address this problem. We show that, by utilizing knowledge about the compilation flow, a verification methodology can be obtained which keeps the respectively occurring intermediate representations close to the identity in an almost perfect fashion. By this, the exponential complexity is frequently reduced to a linear or close-to-linear complexity---substantially reducing the runtime of the verification process.
In order to showcase the possible improvements, we consider the compilation flow as it is currently conducted by IBM's Qiskit.\footnote{However, the methods proposed in this work can be tailored to any other compilation flow as well.} 
Experimental evaluations show that, using the proposed method, circuits composed of tens of thousands of operations can be verified within seconds or even less, whereas state-of-the-art techniques frequently time-out or require substantially more runtime.
%In this paper, we show that this potential can be unleashed by additionally utilizing knowledge about the compilation flow itself. 
%To this end, a large portion of this paper deals with carefully reviewing the compilation flow as it is conducted by IBM's Qiskit\footnote{Qiskit was merely chosen as a representative. Similar results can be achieved for other compilation flows.}. 
%This is needed, in order to engineer an extremely efficient strategy tailored towards verifying compilation results.
%As confirmed by experimental evaluations, the resulting scheme allows to verify circuits comprised of tens of thousands of gates within seconds, whereas state-of the art techniques frequently time-out or require substantial runtime.
A corresponding open source implementation of the proposed method is publicly available at \url{https://github.com/iic-jku/qcec}.

The rest of this work is structured as follows: In Section~\ref{sec:background}, we review the necessary basics of quantum circuits and the IBM Q systems needed to keep this work self-contained. 
Then, Section~\ref{sec:motivation} reviews the compilation flow as it is conducted by IBM's Qiskit and elaborates on verifying results of this flow. 
Based on that, Section~\ref{sec:proposed} illustrates how all different aspects are exploited and consequently orchestrated to form a dedicated verification strategy.
The strategy's performance is then evaluated in Section~\ref{sec:experiments}, before Section~\ref{sec:conclusions} concludes the paper.

\section{Quantum Computing and the IBM Q Systems}\label{sec:background}
In this section, we briefly review the concepts of quantum computing~\cite{nielsenQuantumComputationQuantum2010} and introduce the 
%corresponding 
notation used in this work. Besides that, we also review the considered platform, i.e., IBM~Q systems~\cite{ibmqIBM}.
For more detailed information, we refer the interested reader to the provided references.

\subsection{Quantum Circuits}\label{sec:qc}

In quantum computing, the main computational unit is the \emph{qubit}. In contrast to classical bits, a single qubit $q$ can be in an arbitrary superposition of the basis states $\ket{0}$ and $\ket{1}$, i.e., 
\[
\ket{q} = \alpha_0 \ket{0} + \alpha_1 \ket{1}
\]
with $\alpha_0,\alpha_1\in\mathbb{C}$ and $\abs{\alpha_0}^2 + \abs{\alpha_1}^2 = 1$.
An $n$-qubit quantum system can be in an arbitrary superposition of the $2^n$ basis states 
\[
\ket{b_{n-1}}\otimes\cdots\otimes\ket{b_0}=\ket{b_{n-1}\dots b_0} = \ket{\sum_{i=0}^{n-1} b_i 2^i}
\]
with~$b_i\in\{0,1\}$, i.e., 
\[
\ket{q}_n = \sum_{i=0}^{2^n-1} \alpha_i \ket{i} \mbox{ with } \alpha_i\in\mathbb{C} \mbox{ and } \sum_{i=0}^{2^n-1} \abs{\alpha_i}^2 = 1.
\]
In the circuit model of quantum computation, qubits are represented by \emph{wires} and are manipulated by quantum operations (\emph{quantum gates}).
Specifically, a quantum circuit $G$ with $m$ gates, operating on $n$ qubits, is denoted by $G=g_0\dots g_{m-1}$, where each $g_i$ represents a quantum gate acting on (a subset of) $n$ qubits.
This is %conveniently 
usually visualized through \emph{quantum circuit diagrams}, where the qubit wires are drawn as horizontal lines, gates are drawn using a variety of symbols, and progression of time is assumed to happen from left to right. %typically progresses from left to right.

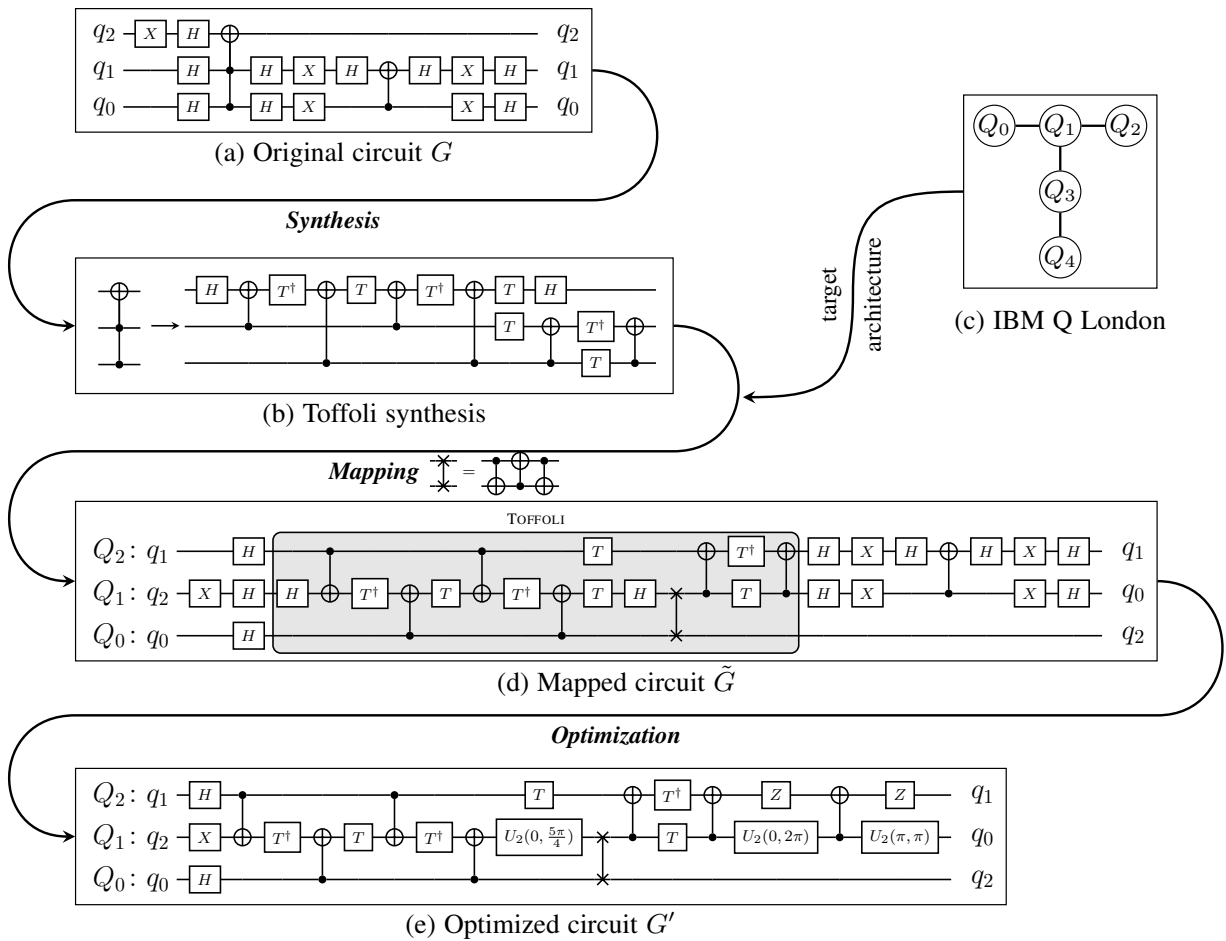
\begin{figure*}[tb]
	\centering
		\resizebox{0.93\linewidth}{!}{
		\begin{tikzpicture}
		\node[draw, rectangle, scale=0.65] (G) {
			\begin{quantikz}[column sep=6pt, row sep={0.65cm,between origins}, ampersand replacement=\&]
			\lstick{\Large$q_2$} \& \gate{X} \& \gate{H} \& \targ{} \& \qw \& \qw \& \qw \& \qw \& \qw \& \qw \& \qw \& \qw \& \rstick{\Large$q_2$}\\
			\lstick{\Large$q_1$} \& \qw \& \gate{H} \& \ctrl{-1} \& \gate{H} \&\gate{X}\& \gate{H} \& \targ{} \& \gate{H} \& \gate{X} \& \gate{H}\& \qw \& \rstick{\Large$q_1$}\\
			\lstick{\Large$q_0$} \& \qw \&\gate{H} \& \ctrl{-2} \& \gate{H} \& \gate{X} \& \qw \& \ctrl{-1} \& \qw \& \gate{X} \& \gate{H}\& \qw \& \rstick{\Large$q_0$}
			\end{quantikz}};

		\node[draw, rectangle, below=2.95cm of G.west, anchor=west, scale = 0.65] (decomp) {
			\begin{tikzpicture}
			\node[] (t) {
				\begin{quantikz}[column sep=6pt, row sep={0.65cm,between origins}, ampersand replacement=\&]
				\& \targ{} \&	\qw\\
				\& \ctrl{-1}\&	\qw\\
				\& \ctrl{-2}\&	\qw
				\end{quantikz}
			};
			
			\node[right= 0.3cm of t] (d) {
				\begin{quantikz}[column sep=6.pt, row sep={0.65cm,between origins}, ampersand replacement=\&]
				\&\gate{H} \& \targ{} \& \gate{T^\dag} \& \targ{} \& \gate{T} \& \targ{} \& \gate{T^\dag} \& \targ{} \& \gate{T} \& \gate{H} \& \qw \& \qw \& \qw\\
				\&\qw \& \ctrl{-1} \& \qw \& \qw \& \qw \& \ctrl{-1} \& \qw \& \qw \& \gate{T} \& \targ{} \& \gate{T^\dag} \& \targ{} \& \qw \\
				\&\qw \& \qw \& \qw \& \ctrl{-2} \& \qw \& \qw \& \qw \& \ctrl{-2} \& \qw \& \ctrl{-1} \& \gate{T} \& \ctrl{-1}									\& \qw
				\end{quantikz}
			};
			
			\draw[thick, -stealth] ($(t.east)+(-0.0,0.00)$) -- ($(d.west)+(0.2,0.00)$);
			\end{tikzpicture}
		};
		
		\node[draw, rectangle, scale=0.65, below=2.95cm of decomp.west, anchor=west] (Gmap) {
			\begin{quantikz}[column sep=6.5pt, row sep={0.75cm,between origins}, ampersand replacement=\&]
				\lstick{\Large$Q_2 \colon q_1$} \& \qw \& \gate{H} \& \qw\gategroup[background,style={fill=gray!20,rounded corners, inner sep=-0.1em},wires=3,steps=14]{{\sc Toffoli}} \& \ctrl{1} \& \qw  \& \qw \& \qw\& \ctrl{1} \& \qw \& \qw \& \gate{T} \& \qw \& \qw \& \targ{} \& \gate{T^\dag} \& \targ{} \& \gate{H} \&\gate{X}\& \gate{H} \& \targ{} \& \gate{H} \& \gate{X} \& \gate{H} \& \qw \& \rstick{\Large$q_1$}\\
				\lstick{\Large$Q_1 \colon q_2$} \& \gate{X} \& \gate{H} \& \gate{H} \& \targ{} \& \gate{T^\dag} \& \targ{} \& \gate{T} \& \targ{} \& \gate{T^\dag} \& \targ{} \& \gate{T} \& \gate{H} \& \swap{1}\&\ctrl{-1} \& \gate{T} \& \ctrl{-1} \& \gate{H} \& \gate{X} \& \qw \& \ctrl{-1} \& \qw \& \gate{X} \& \gate{H} \& \qw \& \rstick{\Large$q_0$}\\
				\lstick{\Large$Q_0 \colon q_0$} \& \qw \& \gate{H} \& \qw \& \qw \& \qw \& \ctrl{-1} \& \qw \& \qw \& \qw \& \ctrl{-1} \& \qw \& \qw \& \targX{} \& \qw \& \qw \& \qw \& \qw \& \qw\& \qw\& \qw\& \qw\& \qw\& \qw\& \qw \& \rstick{\Large$q_2$}
		\end{quantikz}};
		
		\node[draw, rectangle, below=2.95cm of Gmap.west, anchor=west, scale=0.65] (Gp) {
			\begin{quantikz}[column sep=6.5pt, row sep={0.75cm,between origins}, ampersand replacement=\&]
				\lstick{\Large$Q_2 \colon q_1$} \& \gate{H} \& \ctrl{1} \& \qw  \& \qw \& \qw\& \ctrl{1} \& \qw \& \qw \& \gate{T} \& \qw \& \targ{} \& \gate{T^\dag} \& \targ{} \& \gate{Z} \& \targ{} \& \gate{Z} \& \qw \& \rstick{\Large$q_1$}\\
				\lstick{\Large$Q_1 \colon q_2$} \& \gate{X} \& \targ{} \& \gate{T^\dag} \& \targ{} \& \gate{T} \& \targ{} \& \gate{T^\dag} \& \targ{} \& \gate{U_2(0,\tfrac{5\pi}{4})} \& \swap{1}\&\ctrl{-1} \& \gate{T} \& \ctrl{-1} \& \gate{U_2(0,2\pi)} \& \ctrl{-1} \& \gate{U_2(\pi, \pi)} \& \qw \& \rstick{\Large$q_0$}\\
				\lstick{\Large$Q_0 \colon q_0$} \& \gate{H} \& \qw \& \qw \& \ctrl{-1} \& \qw \& \qw \& \qw \& \ctrl{-1} \& \qw \& \targX{} \& \qw\& \qw\& \qw\& \qw\& \qw\& \qw\& \qw \& \rstick{\Large$q_2$}
		\end{quantikz}};
		
		\node[draw, rectangle, scale=0.95, above= 4.5cm of Gmap.east, anchor=east] (arch) {
			\begin{tikzpicture}
			\node[vertex] (0)  at (0,1.6) {$Q_0$};
			\node[vertex] (1)  at (0.8,1.6) {$Q_1$};
			\node[vertex] (2)  at (1.6,1.6) {$Q_2$};
			\node[vertex] (3)  at (0.8,0.8) {$Q_3$};
			\node[vertex] (4)  at (0.8,0) {$Q_4$};
			
			\draw[>=stealth, thick] (0) -- (1);
			\draw[>=stealth, thick] (2) -- (1);
			\draw[>=stealth, thick] (3) -- (1);
			\draw[>=stealth, thick] (3) -- (4);
			\end{tikzpicture}
		};
		
		\node[inner sep=0pt,text width=0.3\linewidth] at ($(G.south)+(0,-0.15)$) (la) {\subcaption{Original circuit $G$}\label{fig:g}};
		\node[inner sep=0pt,text width=0.3\linewidth] at ($(decomp.south)+(0,-0.15)$) (lb) {\subcaption{Toffoli synthesis}\label{fig:decomp}};
		\node[inner sep=0pt,below=0.15\belowcaptionskip of arch,text width=0.2\linewidth] (lc) {\subcaption{IBM Q London}\label{fig:arch}};
		\node[inner sep=0pt,text width=0.3\linewidth] at ($(Gmap.south)+(0,-0.15)$) (ld) {\subcaption{Mapped circuit $\tilde{G}$}\label{fig:gmap}};
		\node[inner sep=0pt,text width=0.3\linewidth] at ($(Gp.south)+(0,-0.15)$) (le) {\subcaption{Optimized circuit $G^\prime$}\label{fig:gp}};

		%\draw[thick,>=stealth,->] (la.south) -- (lb.north) node[midway, left] (synthesis) {\bfseries\emph{Synthesis}};
		%\draw[thick,>=stealth,->] (lb.south) -- (ld.north) node[midway, left] (mapping) {\bfseries\emph{Mapping}};
		%\draw[] (lb.south) -- (ld.north) node[midway, right] (swap) {$\left(\scalebox{0.5}{\begin{quantikz}[column sep=0.1cm, row sep={0.6cm,between origins}, ampersand replacement=\&, inner sep=0pt, outer sep=0pt] 
		%		\& \swap{2} \& \qw \&\&\& \ctrl{2} \& \targ{} \& \ctrl{2} \& \qw \\[-0.35cm] \&\&\&=\&\\[-0.35cm]
		%		\& \targX{} \& \qw \&\&\& \targ{} \& \ctrl{-2} \& \targ{} \& \qw
		%		\end{quantikz}}\right)$};
		%\draw[thick,>=stealth, ->] (ld.south) -- (le.north) node[midway,left] {\bfseries\emph{Optimization}};
		
		\draw[thick,>=stealth,->] (G.east).. controls +(0:1) and +(0:1) ..
		($(G.east)+(0,-1.5)$) .. controls +(180:1) and +(0:1) ..
		($(decomp.west)+(0,1.45)$) node[midway, below] {\bfseries\emph{Synthesis}} 
		.. controls +(180:1) and +(180:1) .. (decomp.west);
		
		\draw[thick,>=stealth,->] (decomp.east).. controls +(0:1) and +(0:1) ..
		($(decomp.east)+(0,-1.45)$) .. controls +(180:1) and +(0:1) ..
		($(Gmap.west)+(0,1.5)$) node[midway, below] (lmap) {\bfseries\emph{Mapping}}
		.. controls +(180:1) and +(180:1) .. (Gmap.west);
		\node[] at ($(lmap.east)+(0.7,-0)$) {$\scalebox{0.65}{\begin{quantikz}[column sep=0.1cm, row sep={0.4cm,between origins}, ampersand replacement=\&, inner sep=0pt] 
				\& \swap{2} \& \qw \&\&\& \ctrl{2} \& \targ{} \& \ctrl{2} \& \qw \\[-0.175cm] \&\&\&=\&\\[-0.175cm]
				\& \targX{} \& \qw \&\&\& \targ{} \& \ctrl{-2} \& \targ{} \& \qw
				\end{quantikz}}$};
		
		\draw[thick,>=stealth,->] (Gmap.east).. controls +(0:1) and +(0:1) ..
		($(Gmap.east)+(0,-1.55)$) .. controls +(180:1) and +(0:1) ..
		($(Gp.west)+(0,1.4)$) node[midway, below] {\bfseries\emph{Optimization}} .. controls +(180:1) and +(180:1) .. (Gp.west);
		
		\draw[thick,>=stealth,->] (arch.west).. controls +(180:1) and +(90:1) ..
		($(arch.west)!0.5!($(decomp.east)+(0.8,-0.825)$)$) node[rotate=90,above] {target} node[rotate=90,below]{architecture} .. controls +(270:1) and +(0:1) ..
		($(decomp.east)+(0.8,-0.825)$);
		
		%\draw[thick,>=stealth, ->] (arch.-160) -- ($($(decomp.east)+(0,-1.7)$)!0.5!(decomp.east)+(0.8,0)$) node[midway,above] {\emph{target}} node[midway,below] {\emph{architecture}};
\end{tikzpicture}}
\caption{Exemplary illustration of the IBM compilation flow}
\label{fig:compilationflow}
\end{figure*}

\begin{example}\label{ex:circuit}
	An example of a quantum circuit $G$ with $16$ gates acting on three qubits %$q_i$ 
	is shown in Fig.~\ref{fig:g}.
	This sequence of operations describes a small instance of the famous \emph{Grover} search algorithm~\cite{groverFastQuantumMechanical1996}.  
	The small boxes with identifiers correspond to operations applied to single qubits such as 
%	The shown circuit contains 
\qop{X} gates (the quantum analogue to the \qop{NOT} gate) and \qop{H}(adamard) gates (which can be used to set a qubit into superposition).  
	Moreover, there are \mbox{\emph{multiple-controlled}~\qop{X}} operations, where an \qop{X} operation is only applied to a \emph{target} qubit (denoted by~$\oplus$) if all of its \emph{control} qubits (denoted by $\bullet$) are in state~$\ket{1}$. In case there is only one control qubit, such a gate is also called \qop{CNOT} or \mbox{controlled-\qop{NOT}}, while in case of two control qubits it is also called a \emph{Toffoli} gate.
\end{example}

Initially, quantum algorithms are described in a way which is agnostic of the device they are planned to be executed on.
However, physical devices today only support very \mbox{low-level} quantum operations. 
In this work, we focus on devices provided by IBM, which are briefly reviewed next.

\subsection{The IBM Q Systems}\label{sec:ibmq}
In 2016, IBM launched the \emph{IBM Quantum Experience} cloud platform which, for the first time, provides public access to a quantum computer.
Today, eight devices with either one, five, or $15$ qubits are freely available for development.
All quantum computers developed at IBM provide the same \textbf{limited \mbox{gate-set}} consisting of arbitrary single-qubit gates~\qop{U} and the two-qubit \qop{CNOT} operation.\footnote{Specifically, $\qop{U}=U_3(\theta,\phi,\lambda)$ with $0\leq \theta \leq \pi$ and $0\leq \phi < 2\pi$ are supported. Special cases are $U_2(\phi,\lambda) = U_3(\frac{\pi}{2}, \phi, \lambda)$ and \mbox{$U_1(\lambda) = U_3(0,0,\lambda)$}.}
Although seemingly small, this already constitutes a universal gate-set, i.e., every possible quantum computation can be realized using only gates from this set~\cite{nielsenQuantumComputationQuantum2010}.
In addition, the IBM Q systems (or in general quantum computers based on superconducting qubits) feature a severely \textbf{limited connectivity} of their qubits. 
This is usually described by a \emph{coupling graph}, where the graph's nodes represent the qubits and an edge between two nodes indicates that a \qop{CNOT} operation may be applied to those qubits\footnote{Nowadays, these edges are typically undirected, i.e., it does not matter which qubit acts as control or target---whereas past architectures actually prescribed the direction.}.
\begin{example}\label{ex:coupling}
The coupling graph of the IBM Q London system is shown in Fig.~\ref{fig:arch}. It consists of five (physical) qubits $Q_i$ and a \qop{CNOT} operation may only be applied to the qubits $(Q_0,Q_1),\,(Q_1,Q_2),\,(Q_1,Q_3)$, or $(Q_3,Q_4)$.
\end{example}
A device's physical qubits are inherently affected by noise---leading to rather \textbf{short coherence times} and \textbf{limited fidelity} of the individual operations.
Until a certain threshold concerning the number of available qubits  is reached, error correction is not yet an option.
%Consequently, today's systems are categorized as \emph{Noise-Intermediate-Scale-Quantum} (NISQ) devices~\cite{preskillQuantumComputingNISQ2018}.
Throughout this work, we will refer to the \emph{physical} qubits of a device using upper case~$Q_i$ and denote the \emph{logical} qubits of a quantum algorithm with lower case~$q_i$. 
Additionally, $Q_i\colon q_j$ denotes %symbolizes 
that the logical qubit~$q_j$ is assigned to the physical qubit~$Q_i$.

\section{Motivation and Considered Problem}\label{sec:motivation}
In order to execute a conceptual quantum algorithm on an actual device such as provided by the IBM Q systems, several transformations have to be applied to the original circuit in order to conform to all the restrictions imposed by the targeted device. 
This transformation process is frequently called \emph{compilation}\footnote{Other terms (interchangeably) referring or relating to (steps of) this process are synthesis, mapping, qubit allocation, qubit routing, or transpilation. In this work, we consider the complete process as \emph{compilation}---split into the three main tasks \emph{synthesis}, \emph{mapping}, and \emph{optimization}.} due to its similarity to a classical compiler transforming high-level code into a (\mbox{machine-executable}) assembly. 
Many specialized tools for this complex task exist, e.g., IBM's Qiskit~\cite{aleksandrowiczQiskitOpensourceFramework2019}, Google's Cirq~\cite{CirqPythonFramework}, Microsoft's QDK~\cite{QuantumDevelopmentKit}, or Rigetti's Forest~\cite{ForestSDK}.
However, to date, the circuits resulting from these tools are hardly verified, i.e., it is hardly checked whether the resulting quantum circuit description still realizes the same functionality as the originally given circuit.
%---posing , which is mainly due to the immense complexity of checking if two quantum circuits indeed realize the same functionality.
In the following, we review this compilation flow using IBM's Qiskit compilation flow as a representative. Afterwards, we elaborate on the problem of verifying results of this flow, which will eventually motivate this work.

\subsection{The IBM Qiskit Compilation Flow}\label{sec:compflow}

%As reviewed in Section~\ref{sec:ibmq}, there are three\todo{nicht explizit eingeführt als 3} 
Compilation of quantum circuits addresses the three kinds of restrictions which limit the usability of a quantum computer and have been reviewed above: The first two, i.e., the limited gate-set and connectivity, constitute \emph{hard} constraints---a computation not conforming to these restrictions may not be executed on the device. In contrast, the %limited 
short coherence time and limited gate fidelity represent %more of 
a \emph{soft} constraint---a quantum circuit may be executed on a device, but it is not guaranteed to produce meaningful results if the circuit, e.g., is too large for the state to stay coherent.
The Qiskit compilation flow is structured as a collection of individual passes, each of which is responsible for dealing with a certain constraint (or an aspect thereof).
Just as traditional compilers, there are different optimization levels offering a trade-off between compilation runtime and quality %performance 
of the compilation result. More precisely:

First, the gates of the original quantum circuit are \textbf{synthesized} to the gate-set supported by the targeted device. 
Most importantly, since devices typically only support up to \mbox{two-qubit} gates, any gate acting on more than two qubits is broken down into ``elementary'' gates. 
This process may require the use of additional \emph{ancillary} qubits for realizing the desired operation. 
In this regard, Qiskit provides several modes, e.g., for the decomposition of multi-controlled gates---offering a \mbox{trade-off} between circuit size and number of required ancillary qubits~\cite{barencoElementaryGatesQuantum1995, maslovAdvantagesUsingRelative2016,willeImprovingMappingReversible2013}.
\begin{example}
	Consider again the circuit $G$ from Ex.~\ref{ex:circuit} as shown in Fig.~\ref{fig:g}. If this circuit shall be executed on an IBM Q system, the Toffoli gate (the two-controlled \qop{NOT}) first has to be realized using  only arbitrary single-qubit gates and~\qop{CNOT}s. One possible synthesized version is shown in Fig.~\ref{fig:decomp}. It takes six \qop{CNOT}s, nine single qubit gates, and no additional ancillaries to realize the desired gate.
\end{example}
Now, the circuit just contains elementary gates supported by the device, but it may not yet conform to the device's limited connectivity. 
Thus, the quantum circuit is \textbf{mapped} to the target architecture, i.e., a mapping between the circuit's \emph{logical} and the device's \emph{physical} qubits is established. Qiskit provides several heuristics for determining an \emph{initial mapping}---from a trivial one-to-one mapping ($Q_i\colon q_i$) to explicitly considering calibration data and picking the most reliable set of qubits for the computation~\cite{muraliNoiseadaptiveCompilerMappings2019}.
However, in most cases, it is not possible to globally define a mapping which conforms to \emph{all} connectivity limitations.
% define an initial mapping so that none of the circuit's gates violates a restriction.
As a consequence, the \mbox{logical-to-physical} qubit mapping usually is %has to 
changed dynamically throughout the circuit.
Typically, this is accomplished by inserting \qop{SWAP} gates into the circuit---effectively allowing to change the mapping of logical qubits to physical qubits so that all operations can be executed while, at the same time, all connectivity constraints are satisfied.
To this end, Qiskit per default uses a very fast, stochastic solution (based on Bravyi's algorithm).
Several other approaches have been proposed for tackling this immensely complex task\footnote{In fact, the mapping task has been shown to be NP-complete~\cite{siraichiQubitAllocation2018}.}~\cite{zulehnerEfficientMethodologyMapping2019, smithQuantumComputationalCompiler2019, willeMappingQuantumCircuits2019, liTacklingQubitMapping2019, matsuoReducingOverheadMapping2019, muraliNoiseadaptiveCompilerMappings2019,liTacklingQubitMapping2019,siraichiQubitAllocation2018,amyStaqFullstackQuantum2019} and some of them have even been integrated into Qiskit.

\begin{example}\label{ex:mapping}
	Consider again the circuit $G$ from Ex.~\ref{ex:circuit} and assume that the Toffoli gate has been synthesized as shown in Fig.~\ref{fig:decomp}. Further, assume that the circuit is to be executed on the IBM Q London architecture shown in Fig.~\ref{fig:arch}. Then, Fig.~\ref{fig:gmap} shows one possible circuit $\tilde{G}$ resulting from this mapping process. The physical qubits $Q_0$, $Q_1$, and $Q_2$ were chosen and initially assigned logical qubits $q_0$, $q_2$, and $q_1$, respectively. Just one \qop{SWAP} operation applied to $Q_0$ and $Q_1$  (indicated by~$\times$) was %needed 
	added in the middle of the circuit in order to conform to the target's connectivity constraints\footnote{A \qop{SWAP} operation is eventually realized using three \qop{CNOT} operations as indicated in the middle of Fig.~\ref{fig:compilationflow}.}.
\end{example}

After this step of the compilation flow, circuits are ready to be executed on the targeted devices (corresponding to the most basic optimization level~O$0$).
However, the previous steps significantly increased the size of these circuits---impacting the achievable performance due to the limited coherence time and gate fidelity.
Thus, several \textbf{optimizations} may be employed to reduce the circuit's size and, hence, improve the actual performance on the quantum computer.
Since the IBM~Q~systems natively support arbitrary single-qubit gates, any number of subsequent single-qubit gates may be \emph{fused} into one single gate.
Additionally, \emph{adjacent-gate-cancellations} can be used to eliminate instances where a gate is directly followed by its inverse, e.g., two consecutive \qop{CNOT} operations with the same control and target qubits can be cancelled.
These are the most basic optimizations that constitute the standard optimization level~O$1$ of Qiskit.
Naturally, more sophisticated optimization techniques have been developed, e.g., gate transformation and commutation~\cite{itokoOptimizationQuantumCircuit2020} (which is included in optimization level O$2$) or re-synthesis of two-qubit unitary blocks~\cite{vidalUniversalQuantumCircuit2004} (which is part of the top optimization level O$3$).

\begin{example}\label{ex:optimized}
	Consider again the circuit $\tilde{G}$ from Ex.~\ref{ex:mapping} shown in Fig.~\ref{fig:gmap} that has been mapped to the IBM Q London architecture. 
	Applying one-qubit-fusion and \mbox{adjacent-gate-cancellation} eventually allows to eliminate nine single-qubit gates and results in the \emph{optimized} circuit $G^\prime$ shown in Fig.~\ref{fig:gp}.
\end{example}

\subsection{Verifying Results of the Compilation Flow}\label{sec:verifyflow}

Naturally, it is of utmost importance that the originally intended functionality of a quantum algorithm is preserved throughout the whole compilation flow.
This can be guaranteed by \emph{verifying} the results of the compilation flow.
To this end, two possible approaches can be employed: (1)~systematically verifying the compilation methods themselves, e.g., by using formal verification techniques~\cite{hietalaVerifiedOptimizerQuantum2019, shiContractbasedVerificationRealistic2019}, or (2)~checking the functional equivalence of the original circuit to the respectively compiled circuit~\cite{viamontesCheckingEquivalenceQuantum2007,smithQuantumComputationalCompiler2019, wangXQDDbasedVerificationMethod2008, yamashitaFastEquivalencecheckingQuantum2010, burgholzerAdvancedEquivalenceChecking2020,niemannEquivalenceCheckingMultilevel2014}.
In this work, we consider the second approach, because, although the first approach allows to guarantee the validity of results for arbitrary circuit inputs, their applicability is severely limited by the effort required to adapt and extend these methods for new developments, such as new optimizations or mapping strategies.

Checking the equivalence between two circuits boils down to checking whether they indeed realize the same functionality.
Mathematically, the quantum gates~$g_i$ of an $n$-qubit quantum circuit~$G$ are defined by $2^n\times 2^n$ unitary matrices~$U_i$.
Consequently, the functionality of a quantum circuit~$G$ with gates $g_0,\dots,g_{m-1}$ is described by a unitary matrix~$U$, which is obtained by consecutively multiplying the unitary matrix representations~$U_i$ of each gate~$g_i$ in reverse order, i.e.,~\mbox{$U=U_{m-1}\cdots U_0$}.
Thus, checking the equivalence of two circuits $G$ and $G^\prime$ amounts to building and comparing the circuits' system matrices $U$ and $U^\prime$.
While simple in its concept, the exponential size of the involved matrices quickly renders many direct approaches %, such as arrays, 
infeasible\footnote{Equivalence checking of quantum circuits has even been proven to be \mbox{\textsf{QMA}-complete} in the general case~\cite{janzingNonidentityCheckQMAcomplete2005}.}.

In order to cope with this complexity, decision diagrams %\emph{Decision Diagrams}~(DDs) 
have been proposed as an efficient data-structure for canonically representing and manipulating quantum functionality in the recent past~\cite{niemannQMDDsEfficientQuantum2016, wangXQDDbasedVerificationMethod2008, zulehnerHowEfficientlyHandle2019}\footnote{Canonicity implies that a comparison of the root pointers of two decision diagrams allows to decide their equivalence.}.
While decision diagrams indeed frequently allow to represent quantum functionality in a very compact fashion, the decision diagrams corresponding to the circuits $G$ and $G^\prime$ may still grow exponentially in the worst case.

Hence, to further improve upon that, a promising approach was recently proposed in~\cite{burgholzerAdvancedEquivalenceChecking2020}  utilizing the following observation:
Consider two equivalent quantum circuits \mbox{$G=g_0\dots g_{m-1}$} and \mbox{$G^\prime =g^\prime_0,\dots,g^\prime_{m^\prime -1}$}. 
Then, due to the inherent reversibility of quantum circuits, this certainly allows for the conclusion that $G^{\prime -1}\cdot G = \mathbb{I}$,  where $G^{\prime -1}$ denotes the inverse of $G^{\prime}$ and  $\mathbb{I}$ denotes the identity function.
Moreover, it holds that
\begin{align*}
\mathbb{I} = G^{\prime -1} \cdot G &=  {(g^{\prime -1}_{m^\prime-1}\dots g^{\prime -1}_0) \cdot (g_0\dots g_{m-1})} \\
&\equiv {(U_{m-1}\cdots U_0)\cdot(U_0^{\prime \dag} \cdots U_{m^\prime -1}^{\prime \dag})} \\
& = {U_{m-1}\cdots U_0\cdot \mathbb{I} \cdot U_0^{\prime \dag} \cdots U_{m^\prime-1}^{\prime \dag}} \\
&\eqqcolon G \shortrightarrow \mathbb{I} \shortleftarrow G^{\prime}.
\end{align*}
As a consequence, checking whether $G$ and $G^\prime$ are in fact equivalent can be conducted by starting with the identity and, then, either applying operations of $G$ ``from the left'' (denoted by $G\shortrightarrow\mathbb{I}$) or (inverted) operations of $G^\prime$ ``from the right'' (denoted by $\mathbb{I}\shortleftarrow G^{\prime}$) until all operations have been applied.
If, afterwards, the identity still remains, the circuits $G$ and $G^\prime$ have been proven to be equivalent.

The intention of this idea is to keep the intermediate computations as close to the identity as possible, since the identity constitutes the best case for most representations of quantum functionality (e.g., linear in the number of nodes with respect to the number of qubits for decision diagrams).

\begin{example}
	Assume, w.l.o.g, that $m\leq m^\prime$, i.e., $G^\prime$ has at least as many gates as $G$.
	Further assume an \emph{oracle} \mbox{$\omega\colon G \to (G^{\prime})^*$} exists that, given a gate $g_i\in G$, returns a consecutive sequence of gates \mbox{$g^\prime_k\dots g^\prime_l\in G^\prime$} such that \mbox{$g_i\equiv g^\prime_k\dots g^\prime_l$}.
	Then, subsequently applying one gate $g\in G$ and $\abs{\omega(g)}$ inverted gates from $G^\prime$ constitutes a ``perfect'' strategy---yielding the identity after each pair of applications.
	As a result, only matrices representing, or staying close to, the identity occur. Since these can usually be represented very efficiently using, e.g., decision diagrams, the process of equivalence checking is substantially improved.
\end{example}
%The authors identify the potential of this equality being the order in which the individual gates from $G$ and $G^{\prime -1}$ are applied---the goal being to stay as close to the identity as possible, since the identity constitutes the best case for most representations of quantum functionality (e.g., linear in the number of nodes with respect to the number of qubits for decision diagrams).
%
%To this end, the proposed methodology starts with a decision diagram representing the identity (i.e, the most compact decision diagram) and, then, subsequently applies gates from $G$ or (inverted) gates from $G^\prime$ according to some strategy, which is symbolized as $G\shortrightarrow\mathbb{I}\shortleftarrow G^\prime$.
%It is stated that, whenever a ``good'' strategy for conducting $G\shortrightarrow\mathbb{I}\shortleftarrow G^\prime$ is available, the whole equivalence check can be conducted very efficiently, since applications of gates from $G$ (potentially increasing the size of the representing DD), can frequently be reverted by applying (inverted) gates from $G^\prime$ (potentially decreasing the size of the representing DD back to that of the identity).

However, a major problem remains in how to obtain the ``perfect'' oracle~$\omega(\cdot)$, i.e., in deciding when to apply operations of $G$ (``from the left'') and when to apply operations of $G^\prime$ (``from the right'').
In~\cite{burgholzerAdvancedEquivalenceChecking2020}, several strategies for this purpose have been proposed and, indeed, substantial speed-ups in checking the equivalence of two quantum circuits have been achieved with that (cf.~Table~1 in~\cite{burgholzerAdvancedEquivalenceChecking2020}).
But after all, those strategies remain rather simple (e.g., they employ a \mbox{one-to-one} or size-proportional application of gates from $G$ and~$G'$) and certainly do not resemble a ``perfect'' strategy which can indeed keep the computation of $G \shortrightarrow \mathbb{I} \shortleftarrow G^{\prime}$ close to the identity.

In contrast, the compilation flow as reviewed in Section~\ref{sec:compflow} provides detailed insights how a circuit~$G$ is eventually compiled to a circuit~$G'$---providing ideal knowledge about how to derive the ``perfect'' oracle~$\omega(\cdot)$.
In this work, we propose a verification scheme which uses the idea of applying $G \shortrightarrow \mathbb{I} \shortleftarrow G^{\prime}$ and, at the same time, utilizes the knowledge about an actual compilation flow (namely the Qiskit flow). As the experimental evaluations (summarized later in Section~\ref{sec:experiments}) confirm, this allows for drastic speed-ups and, eventually, makes equivalence checking feasible on a large scale.% for practical applications. 

\section{Proposed Verification Scheme}\label{sec:proposed}

In this section, we propose a verification scheme which rests on the ideas discussed above, but additionally utilizes knowledge about the Qiskit quantum circuit compilation flow to derive a much better oracle~$\omega(\cdot)$.
For each step in the compilation flow %(cf.~Fig.~\ref{fig:compilationflow}), 
(i.e., for synthesis, mapping, and optimization), a corresponding strategy for~$\omega(\cdot)$ is derived which keeps applying $G\shortrightarrow\mathbb{I}\shortleftarrow G^\prime$ close to the identity.
Those strategies are described in the following subsections. Afterwards, they are combined to an overall scheme.

%In the following, $G$ identifies the original quantum circuit while $G^\prime$ represents the circuit resulting from the compilation flow.
%Just as the discussions in Section~\ref{sec:compflow}, this section is structured according to the different steps of the compilation flow.

%\subsection{The Effect of Synthesis}\label{sec:decomp}
\subsection{Utilizing Knowledge about the Synthesis Step}\label{sec:decomp}
Considering the first step of the compilation flow, two issues become relevant for determining the ``perfect'' $G\shortrightarrow\mathbb{I}\shortleftarrow G^\prime$ strategy: (1)~each gate $g\in G$ is compiled to a sequence of gates \mbox{$g^\prime_k\dots g^\prime_l\in G^\prime$} and (2)~the circuits $G$ and $G^\prime$ may operate on different numbers of qubits due to the addition of ancillary qubits required for the synthesis.

For the first issue, it can be exploited that the actual decomposition scheme, i.e., into how many elementary gates each of the original circuit's gates is decomposed, is known a priori.
Thus, an \emph{oracle} $\omega(\cdot)$ which, given a gate~$g\in G$, returns the corresponding sequence of gates \mbox{$g^\prime_k\dots g^\prime_l\in G^\prime$}, is explicitly known in this case.
%This can be formalized as a \textbf{cost function} $\bm{f(\cdot)}$ that takes a gate~$g$ and returns the number of resulting elementary gates~$f(g)$. 
Assuming that $G^\prime$ resulted from the synthesis of a given quantum circuit $G$, applying \emph{one} gate from $G$ and $\abs{\omega(g)}$ inverted gates from $G^\prime$ constitutes an optimal strategy for conducting $G\shortrightarrow\mathbb{I}\shortleftarrow G^\prime$---yielding the identity after each step.

\begin{example}\label{ex:f}
	Consider the original circuit $G$ shown in Fig.~\ref{fig:g}. As indicated by Fig.~\ref{fig:decomp}, the Toffoli gate of $G$ needs to be decomposed into elementary gates supported by the architecture, while all other gates of $G$ are already supported. Thus,~$\abs{\omega(g)} = 1$ holds for all $g\in G$ except for the Toffoli gate, where~$\abs{\omega(g)}=15$ holds.
\end{example}

In case both circuits do not operate on the same number of qubits, the corresponding unitaries have different dimensions and cannot be applied directly.
Unfortunately, it is not sufficient to match the qubit count of $G^\prime$ by just augmenting the original circuit with idle qubits.
Since ancillary qubits are always initialized in a particular state (typically~\ket{0}), this leaves some degree of freedom in the overall unitary representation $U^\prime$.
In order to compensate for this degree of freedom, the eventually resulting matrix $U^\prime$ has to be modified as shown in the following example.

	\begin{example}\label{ex:ancillary}
	Consider a unitary $2^n\times 2^n$ matrix $U$ and assume that, w.l.o.g., the last qubit $q_{n-1}$ acts as an ancillary qubit initialized to $\ket{0}$.
	In general, the action of $U$ depending on the state of $q_{n-1}$ is described by the four $2^{n-1}\times 2^{n-1}$ \mbox{sub-matrices}~$U_{ij}$ as illustrated in Fig.~\ref{fig:ancu}. 
	Since the ancillary is initialized to~$\ket{0}$, the sub-matrices corresponding to the transformation from~$\ket{1}$ can be ignored---resulting in the \emph{modified matrix}~${\tilde{U}}$ shown in Fig.~\ref{fig:adapted}.
\end{example}

\begin{figure}[tbp]
	\centering
	\begin{subfigure}[b]{0.495\linewidth}
		\centering
		\resizebox{\linewidth}{!}{
	\begin{tikzpicture}
	\node (U) {$
		\begin{bNiceArray}{C|C}[create-medium-nodes, margin, columns-width = auto, code-after = {
			%\tikz \node [rectangle, fill=red!10, blend mode = multiply, rounded corners = 0.5 mm, inner sep=1pt, fit = (1-1-medium)] {} ;
			%\tikz \node [rectangle, fill=yellow!10, blend mode = multiply, rounded corners = 0.5 mm, inner sep=1pt, fit = (1-2-medium)] {} ;
			%\tikz \node [rectangle, fill=blue!10, blend mode = multiply, rounded corners = 0.5 mm, inner sep=1pt, fit = (2-1-medium)] {} ;
			%\tikz \node [rectangle, fill=green!10, blend mode = multiply, rounded corners = 0.5 mm, inner sep=1pt, fit = (2-2-medium)] {} ;
		}, first-row, first-col, code-for-first-row = \color{gray}\scriptstyle , code-for-first-col = \color{gray}\scriptstyle]
		& \ket{0} & \ket{1} \\
		\ket{0} & U_{00} & U_{10}  \\\hline
		\ket{1} & U_{01} & U_{11}
		\end{bNiceArray}$};
	
	\node (Uname) at ($(U)+(-1.6,-0.1)$) {$U\colon$};
	\node[rotate=90, font=\scriptsize] (from) at ($(U)+(-1.25,-0.15)$) {\color{gray}{to}};
	\node[font=\scriptsize] (to) at ($(U)+(0.25,.6)$) {\color{gray}{from}};
	\node[font=\scriptsize, rotate=35] (to) at ($(U)+(-0.8,.5)$) {\color{gray}{$q_{n-1}$}};
	\end{tikzpicture} }
	\caption{Original matrix $U$}
	\label{fig:ancu}
	\end{subfigure}\hfill
	\begin{subfigure}[b]{0.495\linewidth}
		\centering
				\resizebox{\linewidth}{!}{
\begin{tikzpicture}
\node (Up)  {$
	\begin{bNiceArray}{C|C}[create-medium-nodes, margin, columns-width = auto, code-after = {
		%\tikz \node [rectangle, fill=red!10, blend mode = multiply, rounded corners = 0.5 mm, inner sep=1pt, fit = (1-1-medium)] {} ;
		%\tikz \node [rectangle, fill=yellow!10, blend mode = multiply, rounded corners = 0.5 mm, inner sep=1pt, fit = (1-2-medium)] {} ;
		%\tikz \node [rectangle, fill=blue!10, blend mode = multiply, rounded corners = 0.5 mm, inner sep=1pt, fit = (2-1-medium)] {} ;
		%\tikz \node [rectangle, fill=green!10, blend mode = multiply, rounded corners = 0.5 mm, inner sep=1pt, fit = (2-2-medium)] {} ;
	}, first-row, first-col, code-for-first-row = \color{gray}\scriptstyle , code-for-first-col = \color{gray}\scriptstyle]
	& \ket{0} & \ket{1} \\
	\ket{0} & U_{00} & 0  \\\hline
	\ket{1} & U_{01} & 0
	\end{bNiceArray}$};
\node (Uname) at ($(Up)+(-1.6,-0.1)$) {$\tilde{U}\colon$};
	\node[rotate=90, font=\scriptsize] (from) at ($(Up)+(-1.25,-0.15)$) {\color{gray}{to}};
	\node[font=\scriptsize] (to) at ($(Up)+(0.25,.6)$) {\color{gray}{from}};
	\node[font=\scriptsize, rotate=35] (to) at ($(Up)+(-0.8,.5)$) {\color{gray}{$q_{n-1}$}};
%\node (Upname) at ($(Up)+(-2.2,-0.1)$) {$\tilde{U}\colon$};
\end{tikzpicture}}
\caption{Modified matrix $\tilde{U}$}
\label{fig:adapted}
\end{subfigure}
	\caption{Handling of ancillary qubits}
	\label{fig:ancillary}
\end{figure}

%\subsection{The Effect of Mapping}\label{sec:mapping}
%\subsection{The Effect of Mapping}\label{sec:mapping}
\subsection{Utilizing Knowledge about the Mapping Step}\label{sec:mapping}
\emph{Mapping} to the targeted architecture establishes a connection between the circuit's logical and the device's physical qubits.
Consequently, while the description of $G$ is expressed in terms of logical qubits $q_0,\dots,q_{n-1}$, the circuit $G^\prime$ operates on (a subset of) the device's physical qubits $Q_0,\dots,Q_{N-1}$.
If a non-trivial initial mapping (i.e., anything but $Q_i\colon q_i$) is employed, this leads to the situation that gates from $G^\prime$, although functionally equivalent, are applied to different qubits than the gates of $G$.
Thus, concluding the equivalence of both circuits is not possible by straight-forwardly using the oracle function~$\omega(\cdot)$. 
%During verification, straight-forward application of the respective circuits' gates according to the function $f(\cdot)$ leads to the situation that gates from $G^\prime$, although functionally equivalent, are applied to different qubits than the gates of $G$.
%Consequently, it is impossible to conclude the equivalence of both circuits.
Instead, a \emph{qubit map} ${m(\cdot)}$  is employed, which stores the mapping between the physical qubits of the circuit $G^\prime$ and the logical qubits of the original circuit~$G$, i.e.,~$m(Q_i)=q_j$ if physical qubit $Q_i$ is initially assigned logical qubit $q_j$.
Whenever a gate from $G^\prime$ is to be applied to a certain physical qubit $Q_i$, this is translated to the corresponding logical qubit $m(Q_i)=q_j$---again allowing to stay close to the identity.

\begin{example}\label{ex:qmap}
	Consider the original circuit $G$ and the mapped circuit $\tilde{G}$ shown in Fig.~\ref{fig:g} and Fig.~\ref{fig:gmap}, respectively. While the \qop{X} gate at the beginning of $G$ is applied to the logical qubit $q_2$, it is applied to the physical qubit $Q_1$ in the circuit $\tilde{G}$. In order to fix this mismatch, the qubit map $m(\cdot)$---mapping \mbox{$Q_0\mapsto q_0$}, \mbox{$Q_1\mapsto q_2$}, and \mbox{$Q_2\mapsto q_1$}---is employed. Consequently, the \qop{X} gate of $\tilde{G}$ is applied to $m(Q_1) = q_2$ which now matches the original gate from $G$ perfectly.
\end{example}

However, as discussed in Section~\ref{sec:compflow}, the \mbox{logical-to-physical} qubit mapping of a compiled circuit in general changes dynamically throughout the circuit in order to satisfy all constraints imposed by the device's coupling map. As a consequence, the potential of using the (static) qubit map $m(\cdot)$ in combination with the oracle function $\omega(\cdot)$ to stay close to the identity is significantly diminished. That is, because the dynamically changed mapping again results in a scenario where gates from $G^\prime$ are applied to different qubits than in the circuit $G$.
%As discussed in Section~\ref{sec:compflow}, just determining an initial mapping is generally not sufficient to satisfy all constraints imposed by the architecture. The insertion of \qop{SWAP} operations allows to move the logical qubits around---dynamically changing the logical-to-physical qubit mapping.
%However, this significantly diminishes the possibility of reverting an application of a gate $g$ from $G$ by $f(g)$ operations from $G^\prime$ (using the qubit map $m(\cdot)$) since the dynamically changed mapping again results in the scenario where gates from $G^\prime$ are applied to different qubits.
Therefore, a perfect verification strategy needs to keep track of the changes in the logical-to-physical qubit mapping caused by \qop{SWAP} operations\footnote{\qop{SWAP}s can be reconstructed from consecutive sequences of three \qop{CNOT}s in $G^\prime$ as indicated in the middle of Fig.~\ref{fig:compilationflow}.} and, accordingly, needs to \emph{update the qubit map}~$m(\cdot)$ throughout the verification procedure.

\begin{example}
	Consider again the scenario of Ex.~\ref{ex:qmap}. If the $G\shortrightarrow\mathbb{I}\shortleftarrow G^\prime$ scheme is carried out using the qubit map $m(\cdot)$ defined there, the result would not represent the identity. That is, because the logical-to-physical qubit mapping is changed in the middle of $\tilde{G}$ by a \qop{SWAP} operation applied to $Q_0$ and~$Q_1$.
	Thus, at that specific point, the qubit map $m(\cdot)$ has to be updated accordingly, i.e., it then has to map \mbox{$Q_0\mapsto q_2$}, \mbox{$Q_1\mapsto q_0$}, and \mbox{$Q_2\mapsto q_1$}.
	Through this dynamic change, the computation of $G\shortrightarrow\mathbb{I}\shortleftarrow G^\prime$ remains %can again remain 
	close to the identity and, eventually, proves the equivalence of both circuits. %yield the identity as a result.
\end{example}

%\subsection{The Effect of Optimization}\label{sec:optimization}
\subsection{Utilizing Knowledge about the Optimization Step}\label{sec:optimization}
If no optimizations were to be applied to the circuit resulting from the synthesis and mapping step (which is equivalent to applying Qiskit's O$0$ optimization level), % from the mapping pass, 
the strategies proposed above allow to conduct $G\shortrightarrow\mathbb{I}\shortleftarrow G^\prime$ in a perfect fashion---yielding the identity after each step.
However, optimizations as discussed in Section~\ref{sec:compflow} further alter the circuit---making it harder to verify the resulting circuit. In the following, we cover how to anticipate the effects of the two most common optimizations employed in Qiskit---\emph{single-qubit gate fusion} and \emph{adjacent gate cancellation} (see Section~\ref{sec:compflow}).

\begin{example}
	Consider again the circuit $\tilde{G}$ shown in Fig.~\ref{fig:gmap}. There, the grey box indicates the gates of $\tilde{G}$ realizing the Toffoli gate of the original circuit $G$ shown in Fig.~\ref{fig:g}. The middle qubit thereby contains a \qop{T} gate, which is directly followed by an \qop{H} gate. Accordingly, in the optimized circuit shown in Fig.~\ref{fig:gp}, these have been merged into a single~\qop{U_2(0,\tfrac{5\pi}{4})} gate. %Thus, $\abs{\omega(g)}$ is now no longer 15, but 14 in case of the Toffoli gate (see Ex.~\ref{ex:f}). \todo{
	Thus, $\abs{\omega(g)}=15$ does no longer hold, but has to be modified to $\abs{\omega(g)}=14$ instead in case of the  Toffoli gate (see Ex.~\ref{ex:f}).
\end{example}

In addition to anticipating fusions within individual gate realizations through adaptations of the oracle function $\omega(\cdot)$, a \emph{pre-processing} pass is conducted which fuses consecutive single-qubit gates where they are present in the original circuit~$G$ (e.g., fusing the \qop{H-X-H} cascade at the end of the circuit~$G$ shown in Fig.~\ref{fig:g} to a single \qop{Z} gate).
%Consider the fusion of subsequent single-qubit gates. Then, fusions within the individual gates' decompositions can easily be anticipated through an \textbf{adjustment of the cost function}~$f(\cdot)$. Furthermore, a \textbf{pre-processing pass} can be conducted that fuses consecutive single-qubit gates where they are present in the original circuit $G$.
However, reductions across multiple gates that were decomposed during synthesis cannot be accounted for in this fashion.
Thus, the formerly constructed \emph{perfect} oracle function $\omega(\cdot)$ becomes \emph{approximate}.

\begin{example}
	Consider again the circuit $\tilde{G}$ shown in Fig.~\ref{fig:gmap} and its optimized variant $G^\prime$ shown in Fig.~\ref{fig:gp}. Then, the cancellation of the two consecutive \qop{H} gates in the beginning of $\tilde{G}$ cannot be anticipated through a straightforward adaptation of $\omega(\cdot)$. However, as also confirmed by the experimental evaluations in Section~\ref{sec:experiments}, $\omega(\cdot)$ remains a suitable approximation for staying close to the identity.
\end{example}

\begin{figure}[tbp]
	\centering
	\resizebox{\linewidth}{!}{
		\begin{tikzpicture}
		\node[scale=1.3] (Gswap) {
			\begin{quantikz}[column sep=0.12cm, row sep={0.85cm,between origins}, ampersand replacement=\&]
			\& \lstick{$q_1$}  \& \swap{1}\& \ctrl{1} \& \qw \&\rstick{$q_0$} \\
			\& \lstick{$q_0$}  \& \targX{}\& \targ{} \&  \qw \& \rstick{$q_1$}
			\end{quantikz}
		}; 
		
		\node[right of=Gswap, xshift=3.0cm, scale=1.3]  (G) {
			\begin{quantikz}[column sep=0.12cm, row sep={0.85cm,between origins}, ampersand replacement=\&]
			\& \lstick{$q_1$}  \& \ctrl{1} \& \targ{} \& \ctrl{1}\& \ctrl{1} \& \qw \&\rstick{$q_0$} \\
			\& \lstick{$q_0$}  \& \targ{} \& \ctrl{-1} \& \targ{}\& \targ{} \&  \qw \&\rstick{$q_1$}
			\end{quantikz}
		}; 
		
		\node[right of=G, xshift=3.0cm, scale=1.3] (Gp) {
			\begin{quantikz}[column sep=0.12cm, row sep={0.85cm,between origins}, ampersand replacement=\&]
			\& \lstick{$q_1$}   \& \ctrl{1}\& \targ{} \& \qw \&\rstick{$q_1$?} \\
			\& \lstick{$q_0$}   \& \targ{}\& \ctrl{-1} \&  \qw \&\rstick{$q_0$?}
			\end{quantikz}
		}; 
		\node[scale=1.3] (e1) at ($(Gswap)!0.45!(G)$) {$\equiv$};
		\node[scale=1.3] (e2) at ($(G)!0.55!(Gp)$) {$\equiv$};
		\end{tikzpicture}}

	\caption{\qop{SWAP} and \qop{CNOT} cancellation}
	\label{fig:cnotcancel}
\end{figure}
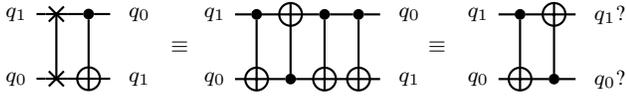

The second optimization employed per default---adjacent gate cancellation---introduces a peculiar issue for the verification strategy as shown in the following example.
\begin{example}
	Consider a \qop{SWAP} operation directly followed by a \qop{CNOT} operation. Since the \qop{SWAP} operation itself is realized by three consecutive \qop{CNOT}s, this sequence of operations may be simplified by cancelling two of them as shown in Fig.~\ref{fig:cnotcancel}. 
	While the qubit map $m(\cdot)$ can be easily adapted in the first two cases, the optimized circuit shows no sign of an applied $\qop{SWAP}$ and, furthermore, introduces an additional \qop{CNOT} gate to the compiled circuit which previously did not exist in $G$. This makes it hard for the proposed strategies to still identify the \qop{SWAP} and, hence, update the qubit map $m(\cdot)$ as described in the previous section.
\end{example}

As a solution, any occurrence of two consecutive \qop{CNOT} operations in $G^\prime$ as shown on the right of Fig.~\ref{fig:cnotcancel} that is not followed by a third matching \qop{CNOT} is \emph{substituted} by the sequence shown on the left of Fig.~\ref{fig:cnotcancel}. Overall, this again allows to accurately track the qubit mapping $m(\cdot)$ and conduct the equivalence check in an optimal fashion.

\subsection{Resulting Verification Scheme}\label{sec:scheme}
All of the considerations above finally result in a dedicated verification scheme that is tailored %-made 
for verifying results of the Qiskit compilation flow.
First of all, a pre-processing step fuses subsequent single-qubit gates in the circuit $G$ and substitutes \qop{SWAP} (and possibly a \qop{CNOT}) gates where applicable in the circuit $G^\prime$.
Afterwards, if necessary, the circuit $G$ is augmented with idle ancillary qubits.
Then, the general $G\shortrightarrow\mathbb{I}\shortleftarrow G^\prime$ scheme %from~\cite{burgholzerAdvancedEquivalenceChecking2020} 
is employed---utilizing the %heuristic 
oracle function $\omega(\cdot)$ to determine which gates from $G^\prime$ are to be applied for each application of a gate from $G$.

The actual application of gates from $G^\prime$ happens with regard to the qubit map $m(\cdot)$ which establishes the connection between the circuit $G$'s logical qubits and $G^\prime$'s physical qubits.
During these steps, this qubit map is dynamically updated to account for the insertion of \qop{SWAP} operations during the mapping.
After applying all gates from both circuits, the result of this scheme is modified as illustrated in Ex.~\ref{ex:ancillary}.
Eventually, the two circuits are shown to be equivalent if the modified result resembles the identity for the non-ancillary qubits.

As confirmed by the experimental evaluations, which are summarized next, this scheme allows to efficiently verify even large instances consisting of tens of thousands of gates within seconds.
Additionally, in contrast to formally verifying the individual compilation steps (see Section~\ref{sec:verifyflow}), this approach remains generic enough to work well out of the box, even when optimizations are employed that have not been directly accounted for, e.g., commutation rules.

\section{Experimental Evaluations}\label{sec:experiments}\vspace{0.2cm}

The proposed verification scheme has been implemented on top of the tool proposed in~\cite{burgholzerAdvancedEquivalenceChecking2020} (which is available at \mbox{\url{https://iic.jku.at/eda/research/quantum_verification}}).
%downloaded from \url{http://iic.jku.at/eda/research/quantum_verification}, which implements the ideas proposed in~\cite{burgholzerAdvancedEquivalenceChecking2020}. 
More precisely, we took the recent version (revision 1.2)
and %In order to realize the scheme proposed in this work (see Section~\ref{sec:scheme}), we 
extended this tool with the strategies described in Section~\ref{sec:proposed}. %\todo{see todo}
%finde diese aufzählung komisch; wir reden im paper davon, dass wir eine perfekte "links/rechts"-strategy entwickeln und dann sagt man hier, dass man unterschiedliche anzahl qubits und mappings unterstützt (du meinst hier schon die Sachen aus IV, aber es geht völlig unter, das damit ja auch die links/rechts-sachen gemeint sind) 
%to (1)~support circuits with different numbers of qubits (see Section~\ref{sec:decomp}), (2)~support logical-to-physical qubit mappings (see Section~\ref{sec:mapping}), and (3)~perform a pre-processing pass for adapting the input circuits (see Section~\ref{sec:optimization}).\todo{die ganze section verbraucht relativ viel platz mit der setup diskussion. das ist mMn aber auch nötig um klarzustellen, was genau evaluiert wurde.}
Afterwards, we conducted extensive experiments to evaluate the performance of the resulting approach.
In this section, we summarize our evaluations. To this end, we first briefly review the setup and, afterwards, present as well as discuss the obtained results.

\subsection{Setup}

In our evaluations, we considered %using 
circuits that are frequently used to benchmark compilers.
%To this end, 
Using IBM Qiskit~\cite{aleksandrowiczQiskitOpensourceFramework2019} (specifically, Qiskit Terra 0.12.0), 
each original circuit $G$ has been compiled for a specific target device---resulting in an alternative circuit $G^\prime$.
During this process, multi-controlled Toffoli gates have been synthesized using the \enquote{\emph{basic}} mode (yielding the smallest circuits, with the highest number of additionally needed qubits) and the target device has been chosen as the smallest possible one capable of accommodating the resulting number of qubits.
Specifically, \emph{IBM Q Boeblingen} has been chosen for circuits with up to 20 qubits, while \emph{IBM Q Rochester} has been used for circuits with up to 53 qubits. 

As discussed in Section~\ref{sec:compflow}, Qiskit offers several optimization levels for compiling circuits. 
In our evaluations, we considered Qiskit's default optimization level (i.e., O$1$) and---in order to show the proposed approach's applicability to scenarios it has not been explicitly tailored towards---the more advanced O$2$ level.
All evaluations have been performed on a \SI{4}{\giga\hertz} \mbox{\emph{Amazon EC2 z1d}} instance running Ubuntu~18.04 with at least \SI{32}{\giga\byte} per job using \mbox{GNU Parallel}~\cite{tangeGNUParallel20182018}. A hard timeout of \SI{1}{\hour} (i.e., \SI[group-minimum-digits = 4]{3600}{\second}) was set for each run.

 \begin{table*}[!t]
	\sisetup{table-text-alignment = right,table-format=>4.2, round-mode = places,round-precision = 2, group-minimum-digits = 4}
	\centering
	\caption{Optimization Level O$1$}
	\label{tab:resultso1}
	\small
	\begin{tabular}{@{}lS[table-format=2.0]S[table-format=5.0]r*{1}{!{\qquad}S[table-format=6.0]SSS}@{}}\toprule
		\multicolumn{4}{c}{Benchmark\hspace*{2.1em}} & \multicolumn{4}{c}{Results}\\
		\cmidrule(r{2.6em}){1-4}\cmidrule(l){5-8}
		{Name} & $n$ & {$\vert G \vert$} & {Architecture} & {$\vert G^\prime\vert$} & {$t_{sota}\,[\si{\second}]$} & {$t_{adv}\,[\si{\second}]$} & {$t_{prop}\,[\si{\second}]$}\\\midrule
				\begin{filecontents*}{./compilationflow_evaluation.csv}
sym9\_317;27;64;53;1540;;63,6108;0,027335;1566;;81,4018;0,032343; 
c2\_182;35;305;53;2545;;2500,79;0,096625;2692;963,183;99,0014;0,221234; 
cm163a\_213;29;39;53;3539;;183,925;0,021553;3704;2004,05;515,692;0,027495; 
apla\_203;22;80;53;13548;;;0,096563;13212;;;0,112028; 
cm151a\_211;28;33;53;3889;;;0,030117;3798;;;0,035532; 
rd84\_313;34;113;53;2295;;;2,63909;2158;;;12,4738; 
mod5adder\_306;32;110;53;2348;;;0,220771;2380;;;0,373354; 
add6\_196;19;229;53;25460;1513,95;514,189;0,505614;26201;455,696;138,647;1,29398; 
cycle17\_3\_112;20;48;53;14987;1111,38;1284,86;0,335453;13876;777,824;667,155;1,80159; 
rd73\_312;25;76;53;1426;1108,62;196,102;0,204738;1498;199,839;226,251;0,648694; 
cm150a\_210;22;53;53;3788;761,817;577,951;0,050704;3678;795,147;583,417;0,068496; 
ryy6\_256;17;44;53;13103;504,418;316,129;0,083114;13100;22,1686;23,3961;0,19964; 
c2\_181;35;116;53;2708;403,797;4,05569;0,028749;3271;;;0,035828; 
alu3\_200;18;94;53;10285;255,254;151,165;0,07072;10463;44,8109;2 7,3967;0,110497; 
decod\_217;21;80;53;6197;192,303;121,174;0,051883;6289;35,5515;23,5979;0,045079; 
example2\_231;16;157;53;19411;187,626;79,6223;0,213097;19207;38,5672;16,7666;0,624709; 
cu\_219;25;40;53;5031;177,145;72,1958;0,024766;4863;136,755;48,825;0,028435; 
plus63mod8192\_164;13;492;53;111720;166,02;112,396;1,85069;111258;110,205;78,3051;8,60594; 
C7552\_205;21;80;53;6384;158,3;80,3795;0,048809;6291;60,1767;32,0745;0,042378; 
alu2\_199;16;157;53;19007;138,998;68,8602;0,172475;19080;23,5608;12,4202;0,268662; 
dk17\_224;21;49;53;6650;105,189;36,5605;0,037176;6589;7,70081;2,99429;0,053159; 
mlp4\_245;16;131;53;14513;79,0592;22,6862;0,094953;14394;12,4061;4,59346;0,165529; 
mux\_246;22;35;53;3679;71,1743;39,7455;0,044385;3829;156,005;91,4915;0,056047; 
urf1\_150;9;1517;20;134216;68,4485;43,9585;9,51565;132898;75,1921;44,026;21,8949; 
clip\_206;14;174;53;20872;62,9079;43,6191;0,201494;21407;44,1643;23,8232;0,256269; 
plus63mod4096\_163;12;429;53;82195;58,4699;44,0082;1,46099;83835;67,0101;45,2387;4,70796; 
dist\_223;13;185;20;19458;19,931;12,0761;0,126195;18721;28,0927;16,6972;0,243709; 
inc\_237;16;93;53;7523;18,0211;8,45923;0,038712;7624;55,0302;42,8347;0,047349; 
urf5\_159;9;499;20;57099;15,9854;9,16781;0,899428;56089;15,5229;8,76742;2,99941; 
alu1\_198;20;32;53;1377;13,1852;4,02241;0,008877;1424;14,5619;1,41585;0,008271; 
cmb\_214;20;18;53;2589;11,4703;10,7618;0,014851;2669;15,0254;15,3057;0,017968; 
sqr6\_259;18;81;53;4436;10,8884;4,50088;0,021385;4536;5,30528;2,02765;0,033187; 
5xp1\_194;17;85;53;5487;10,7763;4,59428;0,036213;5420;6,89312;3,51703;0,043368; 
rd84\_253;12;111;20;7857;9,98725;6,78412;0,05101;7715;4,48325;2,77473;0,055574; 
root\_255;13;99;20;8594;7,96519;5,15064;0,047981;8729;6,1006;3,82038;0,056812; 
\end{filecontents*}
\csvreader[column count=12, no head, separator=semicolon]{./compilationflow_evaluation.csv}
				{1=\name, 2=\qubitsg, 3=\ng, 4=\qubitsgp, 5=\ngpa, 6=\tsotaa, 7=\tadva, 8=\tpropa, 9=\ngpb, 10=\tsotab, 11=\tadvb, 12=\tpropb}
				{\name &
					\qubitsg &
					\ng &
					{\ifthenelse{\lengthtest{\qubitsgp pt < 21 pt}}{Boeblingen}{Rochester}} & 
					\ngpa & 
					{\ifthenelse{\equal{\tsotaa}{}}{\num{>3600.00}}{\num{\tsotaa}}} &
					{\ifthenelse{\equal{\tadva}{}}{\num{>3600.00}}{\num{\tadva}}} &  
					{\ifthenelse{\equal{\tpropa}{}}{\num{>3600.00}}{\ifthenelse{\equal{\tsotaa}{}\or\lengthtest{\tpropa pt < \tsotaa pt}}{\bfseries\num{\tpropa}}{\num{\tpropa}}}} \cr}
				\\[-\normalbaselineskip]\bottomrule
	\end{tabular}\\\vspace{3mm}
{\footnotesize $n$: Number of qubits \hspace*{0.4cm} \emph{$\vert G \vert$}: Gate count of $G$ \hspace*{0.4cm} \emph{Architecture}: Boeblingen (20 qubits) or Rochester (53 qubits) \\ \emph{$\vert G^\prime \vert$}: Gate count of $G^\prime$ \hspace*{0.4cm} $t_{\mathit{sota}}$: Runtime of state-of-the-art EC routine~\cite{niemannQMDDsEfficientQuantum2016} \\ $t_{\mathit{adv}}$: Runtime of advanced methodology EC routine~\cite{burgholzerAdvancedEquivalenceChecking2020} \hspace*{0.45cm}$t_{\mathit{prop}}$: Runtime of proposed, dedicated EC scheme}
\end{table*}

\vspace{-2mm}\subsection{Obtained Results}\vspace{-1mm}

In a first series of evaluations, we considered Qiskit's default optimization level O$1$.
A subset of the respectively obtained results is shown in Table~\ref{tab:resultso1}\footnote{Due to space limitations, only a small subset of benchmarks is listed here. However, the proposed scheme is publicly available at \url{https://github.com/iic-jku/qcec} to conduct further evaluations.}.
 Here, the first columns describe the original circuit~$G$ (its name, number of qubits, and number of gates) as well as the device the circuit was mapped to.
 Then, the size of the resulting circuit~$G^\prime$ is listed, along with the runtimes of (1)~the equivalence checking routine from~\cite{niemannQMDDsEfficientQuantum2016}, 
 (2)~its advanced improvement from~\cite{burgholzerAdvancedEquivalenceChecking2020} (utilizing $G \shortrightarrow \mathbb{I} \shortleftarrow G^{\prime}$)\footnote{\resizebox{0.95\linewidth}{!}{The \enquote{\emph{Proportional}} strategy was used, as it produced the best results in~\cite{burgholzerAdvancedEquivalenceChecking2020}.}}, and
 (3)~the strategy proposed in this work. 
 
 \begin{table*}[!t]
	\sisetup{table-text-alignment = right,table-format=>4.2, round-mode = places,round-precision = 2, group-minimum-digits = 4}
	\centering
	\caption{Optimization Level O$2$}
	\label{tab:resultso2}
	\small
	\begin{tabular}{@{}lS[table-format=2.0]S[table-format=5.0]r*{1}{!{\qquad}S[table-format=6.0]SSS}@{}}\toprule
		\multicolumn{4}{c}{Benchmark\hspace*{2.1em}} & \multicolumn{4}{c}{Results}\\
		\cmidrule(r{2.6em}){1-4}\cmidrule(l){5-8}
		{Name} & $n$ & {$\vert G \vert$} & {Architecture} & {$\vert G^\prime\vert$} & {$t_{sota}\,[\si{\second}]$} & {$t_{adv}\,[\si{\second}]$} & {$t_{prop}\,[\si{\second}]$}\\\midrule
				\csvreader[column count=12, no head, separator=semicolon]{./compilationflow_evaluation.csv}
				{1=\name, 2=\qubitsg, 3=\ng, 4=\qubitsgp, 5=\ngpa, 6=\tsotaa, 7=\tadva, 8=\tpropa, 9=\ngpb, 10=\tsotab, 11=\tadvb, 12=\tpropb}
				{\name &
					\qubitsg &
					\ng &
					{\ifthenelse{\lengthtest{\qubitsgp pt < 21 pt}}{Boeblingen}{Rochester}} & 
					\ngpb & 
					{\ifthenelse{\equal{\tsotab}{}}{\num{>3600.00}}{\num{\tsotab}}} & 
					{\ifthenelse{\equal{\tadvb}{}}{\num{>3600.00}}{\num{\tadvb}}} &  
					{\ifthenelse{\equal{\tpropb}{}}{\num{>3600.00}}{\ifthenelse{\equal{\tsotab}{}\or\lengthtest{\tpropb pt < \tsotab pt}}{\bfseries\num{\tpropb}}{\num{\tpropb}}}} \cr}		
				\\[-\normalbaselineskip]\bottomrule
	\end{tabular}\\\vspace{2mm}
{\footnotesize $n$: Number of qubits \hspace*{0.4cm} \emph{$\vert G \vert$}: Gate count of $G$ \hspace*{0.4cm} \emph{Architecture}: Boeblingen (20 qubits) or Rochester (53 qubits) \\ \emph{$\vert G^\prime \vert$}: Gate count of $G^\prime$ \hspace*{0.4cm} $t_{\mathit{sota}}$: Runtime of state-of-the-art EC routine~\cite{niemannQMDDsEfficientQuantum2016} \\ $t_{\mathit{adv}}$: Runtime of advanced methodology EC routine~\cite{burgholzerAdvancedEquivalenceChecking2020} \hspace*{0.45cm}$t_{\mathit{prop}}$: Runtime of proposed, dedicated EC scheme}
\end{table*}
 
 The results clearly show the superiority of the proposed method. 
 While the first approach using $G \shortrightarrow \mathbb{I} \shortleftarrow G^{\prime}$ (as proposed in~\cite{burgholzerAdvancedEquivalenceChecking2020})
 allows to reduce the equivalence checking runtime down to a third or a half in most cases (sometimes even more), substantial runtimes (or even timeouts) are still reported.
 On the contrary, additionally exploiting explicit knowledge about the compilation flow as proposed in this work allows for drastic further improvements. In fact, 
% advanced methodology proposed in~\cite{burgholzerAdvancedEquivalenceChecking2020} frequently allows to reduce the equivalence checking time down to a third or a half (sometimes even more), the potential of the \mbox{$G\shortrightarrow\mathbb{I}\shortleftarrow G^\prime$} scheme becomes evident when employing a strategy carefully designed to verify the results of the specific task at hand, i.e., compilation.
 \emph{all} considered instances are successfully verified within (fractions of) seconds, whereas state-of-the-art equivalence checking methods and even the recently proposed advanced techniques frequently time-out or require substantial runtime.

In a second series of evaluations, we considered Qiskit's more advanced optimization level O$2$.
In this regard, Table~\ref{tab:resultso2} shows a representative subset of the obtained results\footnote{In some instances, the circuit resulting from optimization level~O$2$ is actually larger than the circuit resulting from~O$1$. This is due to the fact that the default \qop{SWAP}-insertion technique employed in Qiskit is stochastic. However, this does not influence the general observations gained from this evaluation, since the actual number of \qop{SWAP}s actually does not influence the performance of the proposed strategy.}.
While the proposed methodology was explicitly tailored for the default optimization level of Qiskit, i.e., O$1$, its performance remains almost on an equally high level in case of verifying circuits compiled with optimization level O$2$---where several more advanced optimization techniques, such as gate commutation rules, are employed, which are not directly accounted for in the proposed scheme. 
 %This becomes evident in the second series of evaluations conducted using Qiskit's O$2$ optimization level.
This shows that even utilization of partial knowledge about the underlying compilation flow is sufficient to drastically improve the verification of the correctness of its result.

\section{Conclusion}\label{sec:conclusions}
In this work, we proposed a dedicated scheme for verifying results of the IBM Qiskit compilation flow. 
To this end, we exploit characteristics unique to quantum computing and explicitly incorporate knowledge about the compilation flow in order to design a strategy that allows to keep the overhead of verifying compilation results minimal.
Experimental evaluations confirm that the proposed strategy consistently allows to verify instances with more than ten-thousand gates within seconds---even if optimizations are employed which are not directly accounted for. Compared to the state of the art, which often requires substantial runtimes or even time-outs in these tasks, this is a drastic improvement. 
The resulting tool is publicly available at \url{https://github.com/iic-jku/qcec} and can easily be adapted to different compilation flows or additional optimizations in the future.

\section*{Acknowledgments}
This work has partially been supported by the LIT Secure and Correct Systems Lab funded by the State of Upper Austria as well as by the BMK, BMDW, and the State of Upper Austria in the frame of the COMET program (managed by the FFG).

\printbibliography
\end{document}